\documentclass[prb,amssymb,superscriptaddress,showpacs,floatfix,a4paper,twocolumn]{revtex4}

\usepackage{graphicx}% Include figure files
\usepackage{dcolumn}% Align table columns on decimal point
\usepackage{bm}% bold math

\begin{document}

\title{Ground-state topology of the Edwards-Anderson $\pm J$ spin glass
model}

\author{F. Rom\'a}
\affiliation{Departamento de F\'{\i}sica, INFAP, Universidad Nacional de
San Luis, CONICET, Chacabuco 917, D5700BWS San Luis, Argentina}
\author{S. Risau-Gusman}
\affiliation{Centro At{\'{o}}mico Bariloche, CONICET, San Carlos de
Bariloche, R8402AGP R\'{\i}o Negro, Argentina }
\author{A. J. Ramirez-Pastor}
\affiliation{Departamento de F\'{\i}sica, INFAP, Universidad Nacional de
San Luis, CONICET, Chacabuco 917, D5700BWS San Luis, Argentina}
\author{F. Nieto}
\affiliation{Departamento de F\'{\i}sica, INFAP, Universidad Nacional de
San Luis, CONICET, Chacabuco 917, D5700BWS San Luis, Argentina}
\author{E. E. Vogel}
\affiliation{Departamento de F\'{\i}sica, Universidad de La
Frontera, Casilla 54-D, Temuco, Chile}

\date{\today}

\begin{abstract}
In the Edwards-Anderson model of spin glasses with a bimodal distribution of bonds, the degeneracy of the ground state allows one to define a structure called backbone, which can be characterized by the rigid lattice (RL),
consisting of the bonds that retain their frustration (or lack of it) in all ground states. In this work we have performed a detailed numerical study of the properties of the RL, both in two-dimensional (2D) and three-dimensional (3D) lattices. Whereas in 3D we find strong evidence for percolation in the thermodynamic limit, in 2D our results indicate that the most probable scenario is that the RL does not percolate.  On the other hand, both in 2D and 3D we find that frustration is very unevenly distributed. Frustration is much lower in the RL than in its complement. Using equilibrium simulations we observe that this property can be found even above the critical temperature. This leads us to propose that the RL should share many properties of ferromagnetic models, an idea that recently has also been proposed in other contexts. We also suggest a preliminary generalization of the definition of backbone for systems with continuous distributions of bonds, and we argue that the study of this structure could be useful for a better understanding of the low temperature phase of those frustrated models.
\end{abstract}

     %Spin glasses and other random magnets

\maketitle

%.....................................................................
\section{\label{S1}INTRODUCTION}

During the last three decades, the study of spin glasses (SGs) has
attracted the interest of several researchers in both experimental
and theoretical groups.  In such magnetic system, the disorder
and frustrations give rise to a complex behavior that it is far from 
being completely understood.  A fundamental problem is to determine the
true nature of the low temperature phase.  With this purpose, the experimental 
and simulations results are commonly analyzed in
the framework of two theories: the replica-symmetry breaking (RSB)
or mean field picture \cite{Parisi1983} and the droplet picture.
\cite{Fisher1986}  While the droplet picture predicts a simple
scenario with only two pure states related each other
by an up-down symmetry, by using the formalism of RSB one finds a
non-trivial phase space broken in many ergodic components and with
an ultrametric topology. In spite of the effort put to solve this
problem, the controversy about the phase-space structure of SGs
remains unresolved.

Recently a different approach \cite{Roma2006,Roma2007a,Roma2007b,Roma2010} has been proposed to analyze the
simulation data of the Edwards-Anderson bimodal (EAB) spin glass
model. \cite{EA}  In the same spirit of the droplet picture, which
focuses on the ground state (GS) and their excitations, in this
approach it is assumed that the GS heterogeneities play a
fundamental role to describe the low-temperature behavior of SG
systems. In the EAB model the fundamental level is degenerate and the spatial
heterogeneities are well characterized by the so-called  {\em
rigid lattice} (RL). \cite{Barahona1982} This
structure is composed by the set of bonds which do not change its
condition (satisfied or frustrated) in all the configurations of the
GS. These bonds are called {\em rigid bonds}.  The remaining ones,
called {\em flexible bonds}, form the {\em flexible lattice}
(FL).  In Ref.~\onlinecite{Roma2007a}, it was shown that in three-dimensional (3D)
lattices the distributions of domain-wall energies are very
different in these two lattices: while the defect energy on the RL
shows a dependence with the system size typical of a highly stable
phase (similar to the 3D ferromagnetic Ising model, but with a
fractal dimension larger than 2), on the FL this quantity shows a
very different behavior, rather like a system in an excited state. The
total defect energy, that is, the sum of these two
contributions, shows a low stability with a small (but
positive) stiffness exponent.

The same idea has been used to analyze the
strong heterogeneities observed in the out-of-equilibrium dynamics
of the EAB model.  For example, in the 3D EAB model the mean
flipping time probability distribution function presents two main
peaks, corresponding to fast and slow degrees of freedom.
\cite{Ricci2000} For the two-dimensional (2D) and 3D EAB model, it has been possible
to show that these slow and fast peaks are related to the
sets of {\em solidary} (S) and {\em non-solidary} (NS) spins,
respectively. \cite{Roma2006,Roma2010} The set S consists of spins which
maintain their relative orientation in all configurations of the
GS (the remaining spins are denoted NS spins). The {\em backbone} of the EAB model is characterized both by the bonds of the RL and by the S spins. In addition, the dynamical heterogeneities in the
violation of the fluctuation-dissipation theorem (FDT) for the 3D
EAB model were studied in the same way. \cite{Roma2007b} Dividing
the system into sets S and NS, numerical simulations show that the
violation of FDT is the result of two components with completely
different behaviors: one that tends to satisfy the FDT relation
(set of NS spins), and another which presents a violation of this
relation similar to coarsening systems (set of S spins).

The concept of backbone is also relevant in computer science, and in particular in the analysis of the $K$-satisfiability
($K$-SAT) problem, a paradigmatic model belonging to the NP-complete class of problems. \cite{Monasson1999,Semerjian2008,Zdeborova2009} In this case one has to assign a binary value to a set of variables so as to satisfy the largest possible number of a given set of clauses. The backbone is defined as the set of variables that take the same values in all optimal assignments. It has been found that the phase transition separating satisfiable from unsatisfiable formulas is characterized by the size of the backbone, which arises as a natural order parameter. \cite{Monasson1999} In addition, the nature of the change in the backbone size at the transition (continuous or discontinuous) can be used to explain the onset of exponential complexity that occurs when going from $2$-SAT, a problem solvable in polynomial time, to $3$-SAT, an NP-complete problem.

These studies show that the backbone plays an important role in
the physical behavior of disorder and frustrated systems. In all the
mentioned cases the separation of the systems in two components
is not trivial: the observables evaluated on the backbone region or their
complement behaves very differently. More surprising is the fact that in the
EAB spin glass model the physics displayed in each one of these components
looks similar to a ferromagnetic and a paramagnetic phase of the Ising model, respectively.
Given this context, we think that an exhaustive study of the backbone structure can give interesting insights
to understand the nature of the low temperature phases in spin glasses.

In this work we have carried out a systematic study of the backbone structure of the 2D and 3D EAB spin glass model. We find that frustration is much lower in the backbone than in its complement. Using equilibrium simulations we observe that this property can be found even above the critical temperature. These findings, together with the results mentioned in the previous paragraphs, lead us to propose that a separate study, in the backbone and its complement, of the different quantities that characterize the system, could lead to a better understanding of the low temperature phase of spin glasses.

The paper is structured as follows. In Sec. \ref{ModAlg} we present
the EAB model and the algorithm used to calculate the RL and the S spins.
Numerical results that characterize the backbone structure are presented in Sec. \ref{NumRes}.
Then, an extensive discussion is given in Sec. \ref{Dis} to show the importance of considering
the GS topology in spin glasses. Finally conclusions are drawn in Sec. \ref{Conc}.

%.....................................................................
\section{\label{ModAlg}MODEL AND ALGORITHM}

We start by considering the Hamiltonian of the EAB model,  \cite{EA}
\begin{equation}
H = - \sum_{( i,j )} J_{ij} \sigma_{i} \sigma_{j}, \label{ham}
\end{equation}
where the sum runs over the nearest neighbors of either a 2D (square)
or 3D (cubic) lattice of linear dimension $L$ and $\sigma_i = \pm 1$ are
$N$ Ising spin variables.  The coupling constants are
independent random variables chosen from a $\pm J$ bimodal
distribution with zero mean and variance one (i.e., $J=1$). Samples (particular
realizations of random bond distribution) in 2D were generated with both
periodic-free boundary conditions (pfbc)
and periodic-periodic boundary conditions (ppbc), while in 3D only
periodic boundary conditions in all directions were used. Because of the fact that the bonds are independent variables, for
relatively large system sizes only configurations with half of the bonds of each sign are
statistically significant. To preserve this feature for small
sizes, we explicitly enforce the constraint
\begin{equation}
\sum_{( i,j )} J_{ij} = \left\{
\begin{array}{cl}
0 & \mbox{for even number of bonds} \\
\pm 1 & \mbox{for odd number of bonds}.
\end{array}
\right.
\label{constraint}
\end{equation}
Specifically, for systems with an odd number of bonds, we enforce the constraint
$\sum_{( i,j )} J_{ij} =1$ for half of the samples and
$\sum_{( i,j )} J_{ij} =-1$ for the other half. \cite{Nota2}

\begin{figure}[t]
\includegraphics[width=\linewidth,clip=true]{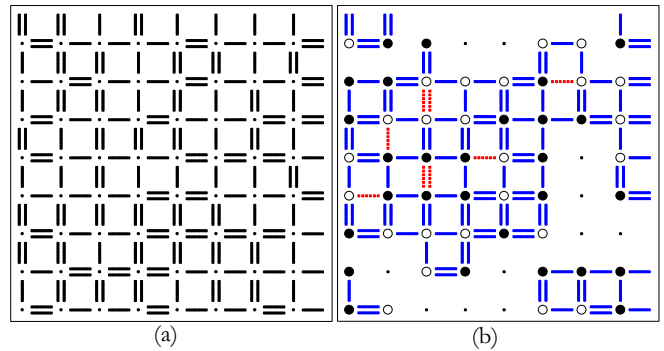}
\caption{(Color online) (a) A typical 2D sample of $L=8$ with ppbc. Single
 and double lines indicate ferromagnetic and anti-ferromagnetic bonds, respectively.
(b) The corresponding RL. Solid (blue) and dot (red) lines
indicate rigid bonds which are, respectively, satisfied and
frustrated in the GS. The S spins are indicated with circles
(closed and open circles correspond to different spin
orientations).} \label{figure1}
\end{figure}

For the EAB model, which has a degenerate GS, the RL is defined as
the set of bonds which do not change its condition (satisfied or frustrated)
through all the configurations of the GS. Figure~\ref{figure1} shows the RL of a typical 2D sample of $L=8$ with ppbc (flexible bonds are not shown).  In 2D, it is only possible to obtain all the GSs only for small system sizes because the number of configurations grows exponentially with $L$. Thus, only for samples with $L \leq 9$ we have performed an exhaustive search of all the GSs, using a branch-and-bound algorithm, in order to obtain the RL. \cite{Hartwig1984} For larger lattice sizes a different approach must be used, because for already $L>9$ the number of GS configurations becomes larger than $10^7$.

Recently we have shown that the RL can be obtained by calculating
only $N_B$ GS configurations, where $N_B$ is the number of bonds. \cite{Ramirez2004} The corresponding algorithm, called rigid lattice searching algorithm (RLSA) is valid for any lattice geometry
in any dimensions. Assuming that one has a method for obtaining a GS configuration of the system, the RLSA can be described as follows:
\begin{enumerate}
\item{For a given sample, a GS configuration $C$ is calculated and its energy $U$ is stored.}

\item{A bond $J_{ij}$ is chosen at random.}

\item{The system being in configuration $C$, one of the spins joined by the bond $J_{ij}$, i.e. either $\sigma_{i}$ or $\sigma_{j}$, is flipped. This changes the ``condition" of the bond from satisfied to frustrated or viceversa.}

\item{Now the orientation of the spins $i$ and $j$ is frozen, and for this ``constrained" system a GS configuration $C^*$ of energy $U^*$ is calculated. The freezing of the spins enforces the constraint that the bond $J_{ij}$ will have the same condition in all GS configurations.}

\item{If $U^* > U$, then $J_{ij}$ is a rigid bond which is
satisfied (frustrated) in all GS configurations of the original, unconstrained, system, if $J_{ij}$
was satisfied (frustrated) in $C$. The bond is added to the RL }

\item{If $U^* = U$, then $J_{ij}$ is a flexible bond, and it is added to the FL.}

\item{The bond $J_{ij}$ is added to the list of ``interrogated" bonds, and the restriction over the spins $\sigma_{i}$ or $\sigma_{j}$ is lifted.}

\item{If there are still non-interrogated bonds, a new bond $J_{ij}$ is chosen among them and the process is repeated from step 3.}
\end{enumerate}

A significant speedup of the algorithm can be obtained if each time that a new GS configuration of the system is obtained (step 6), new configurations are explored by performing single spin flips. There are usually some spins, called {\em free spins}, whose flipping does not change the energy of the state, leading thus to new GSs. This procedure, called {\em invasion}, allows us to explore a local ensemble of GSs (LEG),\cite{Vogel1999} that are related by single spin flips. By comparing the GSs thus obtained many flexible bonds can be detected. Note that this procedure gives no information about rigid bonds, because these must keep their condition in {\em all} GSs. Although the speedup given by this procedure is important, the number of GSs required to obtain the RL remains of the order of $N_B$.

For the procedure presented above we have assumed the existence of an algorithm that can find any GS configuration of the systems involved. But for some systems, if the sample size is not very small, only probabilistic algorithms are available, i.e. algorithms whose output is a GS configuration with a probability smaller than $1$. In this case the only modification of the RLSA is that, in step 5, a bond is classified as rigid if in $n$ independent runs of the probabilistic algorithm the condition $U^{*}> U$ is obtained. Note that, as in this case flexible bonds are always correctly classified, the true RL is a subset of the RL calculated with our algorithm. For lattices with ppbc we implemented the RLSA using parallel tempering Monte Carlo. \cite{Geyer1991,Hukushima1996} Recently, we have shown that this technique is a powerful heuristic method
for reaching the GS of the EAB model up to $L=30$ in 2D and $L=14$ in 3D. \cite{Roma2009} As in the RLSA many independent runs of parallel tempering are needed, we have obtained the RL of samples up to $L=18$ in 2D and $L=9$ in 3D. Parameters used here are the same as in Ref.~\onlinecite{Roma2009} and $n=10$ independent runs were carried out in all the cases.

For 2D lattices with pfbc it is well known that the problem of finding the GS can be mapped to a minimum-weighted
perfect matching problem, which can be solved exactly in polynomial time (i.e in time proportional to some power of $L$).\cite{Hartmannlibro} We have used one implementation of the Blossom algorithm, \cite{Blossom} which has allowed us to obtain the GS of systems with sizes up to $L=300$. To obtain the RL, however, we have not used the RLSA, because systems with frozen spins cannot be mapped to perfect matching problems.  Then, we have implemented an equivalent algorithm that it is roughly as follows:
\begin{enumerate}
\item{A GS configuration $C$ of the system is calculated and its energy $U$ is stored. All bonds are listed as unclassified.}
\item{An unclassified bond $J_{ij}$ is chosen.}
\item{If bond $J_{ij}$ is satisfied, we do $J_{ij} \to -J_{ij}$, and if it is frustrated, we do $J_{ij} \to 2 J_{ij}$. Note that this increases the energy of the configurations where the bond is in the same condition as in $C$, and decreases the
energies of the rest.}
\item{The GS is calculated for the modified system and its energy $U^*$ stored.}
\item{If $U^*<U$, bond $J_{ij}$ is classified as flexible, else it is
classified as rigid}.
\item{Steps 2 to 5 are repeated until all the bonds have been classified.}
\end{enumerate}
The procedure, implemented with the Blossom algorithm, has allowed us to obtain the RL of samples with system sizes up to $L=140$.

This algorithm is based on the following reasoning. If we want to find the rigidity of bond $J_{ij}$, the set of all $2^N$ possible spin configurations can be split into two subsets: set $\mathcal{A}_F$, containing all the configurations where this bond is frustrated, and set $\mathcal{A}_S$, containing all configurations where this bond is satisfied. The energy of the lowest configurations within each set are denoted $U_F$ and $U_S$, respectively. Let us suppose now that the GS configuration $C$ of the system (with energy $U$) belongs to set $\mathcal{A}_F$ and $U=U_F$. Then, we modify the bond $J_{ij}$ in such a way (see below) that for all configurations of $\mathcal{A}_F$ the energy is raised by the same quantity $\delta$. Thus, the lowest energy configurations of this set are the same as in the original system, but with energy $U_F^*=U_F+\delta$.  In turn, the lowest energy of the configurations belonging to set $\mathcal{A}_S$ is lowered by the amount $\delta$, i. e. $U_S^*=U_S-\delta$.

Now, if bond $J_{ij}$ is flexible, we have $U=U_F=U_S$ and therefore the GS energy of the modified system will satisfy $U^*=U_S-\delta=U-\delta$.  On the other hand, if bond $J_{ij}$ is rigid, we have $U_F<U_S$. In this case, the new GS energy is either $U^*=U_S-\delta$ or $U^*=U_F+\delta$. But note that in any of these cases, we have $U^*>U-\delta$. Thus, to know whether a bond is rigid or not it is enough to check if the GS energy of the modified system satisfies $U^*=U-\delta$ (flexible bond) or $U^*>U-\delta$ (rigid bond). 

If the GS configuration $C$ belongs to set $\mathcal{A}_S$, the preceding reasoning is exactly the same. The only feature that depends on whether the GS configuration belongs to set $\mathcal{A}_S$ or $\mathcal{A}_F$, is the modification of the bond to produce a new system with an associated value of $\delta$.  Note, however, that the method works for any value of $\delta$. The values we have chosen are as follows: if the GS configuration belongs to set $\mathcal{A}_F$ we do $J_{ij} \to 2 J_{ij}$, resulting in $\delta=1$, whereas if the GS configuration belongs to set $\mathcal{A}_S$ we do $J_{ij} \to -J_{ij}$, resulting in $\delta=2$. As the smallest energy gap between consecutive levels is $2J$, these choices of $\delta$ guarantee that if the bond is rigid then the new GS energy will be larger than or equal to the GS energy of the original system, and if it is flexible it will be strictly smaller. Thus, as stated in step 5 of the algorithm above, a comparison between the GS energies of the modified and original systems is enough to determine whether the bond is flexible or not.

%.....................................................................
\section{\label{NumRes} NUMERICAL RESULTS}

In this section we present the main topological characteristics of the
backbone for 2D and 3D lattices. First, we analyze the distribution of
backbone sizes and, by extrapolating, we calculate the fraction of rigid
bonds and S spins in the thermodynamic limit. Next, the percolation process
displayed for finite lattice sizes is studied in detail to infer
the structure of the backbone of macroscopic samples. We conclude by showing
the behavior of the contributions of the RL and FL to the mean energy per bond.

\subsection{\label{S3.1}Solidary spins}

\begin{figure}[t]
\includegraphics[width=\linewidth,clip=true]{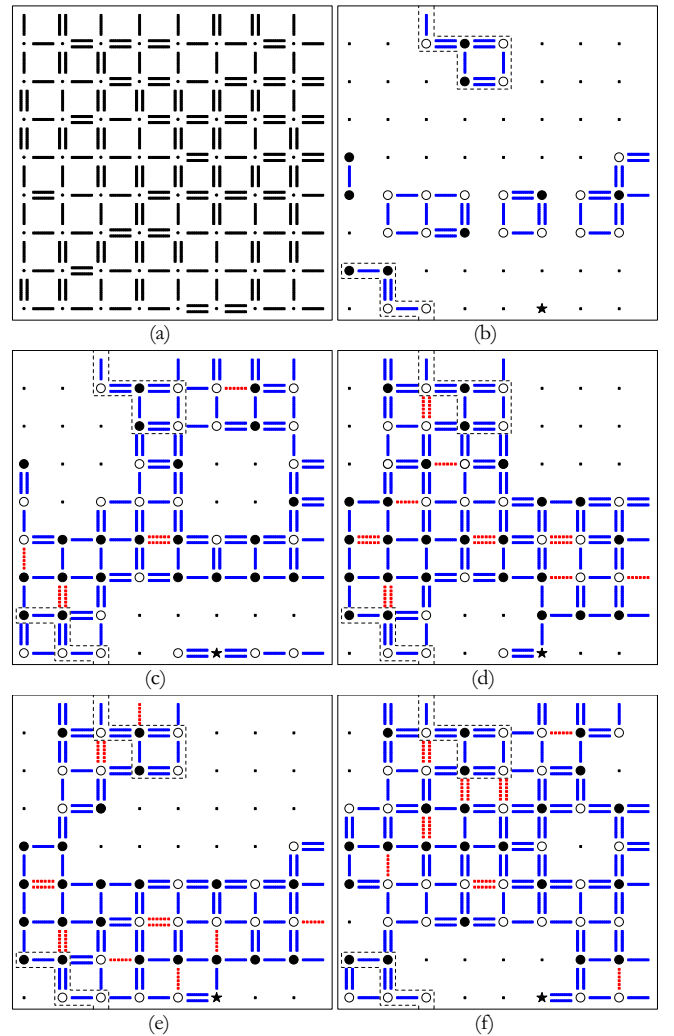}
\caption{\label{figure2} (Color online) (a) A 2D sample of $L=8$ with ppbc. (b)
The corresponding RL formed by four islands.  Symbols are as in 
Fig.~\ref{figure1}. The spin denoted by a star is solidary with
the set of spins contained inside the closed dash line. (c-f) Partial
RLs for the four LEGs in which are broken all configurations of the GS.}
\end{figure}

As mentioned above, the backbone can be characterized by the set of S spins which preserve their relative orientations in all configurations of the GS. It is evident that the spins joined by bonds of the RL are S spins. But, remarkably, there are samples where some solidary spins are ``isolated'', in the sense that they are not connected to the RL. To see how this is possible, let us consider the sample shown in Fig.~\ref{figure2} (a). We have obtained its RL, as well as the set of all solidary spins, by exploring all the GSs of the system, using a branch-and-bound algorithm. Figure~\ref{figure2} (b) shows that the RL is composed by four islands. One of the solidary spins, denoted by a star in Fig.~\ref{figure2} (b), is outside of RL. As mentioned in the preceding section, the set of GS configurations can be divided in several disjoint subsets called LEGs.  A LEG is defined by the requirement that their configurations are connected by paths of single spin-flips.  Within each LEG, the bonds that do not change their status form a ``partial'' RL. The intersection of the partial rigid lattices gives rise to the ``global'' RL. Fig.~\ref{figure2} shows the partial RL of the four possible LEGs for that sample. Note that the spin denoted with a star belongs to the four partial RL. In other words, it is connected by four different paths to the global RL island identified with a dashed line, and is therefore ``solidary'' with it.

Fortunately, analyzing many samples up to sizes of $L=8$ in 2D and $L=4$ in 3D, we have concluded that
the situation exemplified in Fig.~\ref{figure2} is very rare and should not be significant for larger lattice sizes: only $~0.1\%$ and $~1\%$ of samples in, respectively, 2D and 3D lattices have isolated solidary spins (and less than 5 of these spins in each case), and we have not found indications that these percentages increase with increasing $L$.   Therefore, in the following we consider as S spins only those that are linked by the RL. This is necessary because our RLSA can only find sets of rigid bonds.

In a recent work, a bond-diluted version of the 3D EAB has been studied close to the bond percolation threshold of the lattice. \cite{Angelini2010} The critical density of non-zero bonds separating the paramagnetic from the SG phase in the GS, was determined by analyzing the correlation function between two external spins put in both ends of the sample.  In the thermodynamic limit, the authors find that these external spins are perfectly correlated (solidary) up to the critical density.  In other words, they change from solidary to non-solidary at the SG threshold.  From this work, however, it is difficult to infer whether the RL also percolates at the same critical density.  Nevertheless, if the RL percolation threshold is lower, the mechanism shown in Fig.~\ref{figure2} would provide an explanation of why the externals spins keep solidary in absence of a percolating rigid structure. It could thus play an important role in this kind of systems.

\subsection{\label{S3.2}Size of the backbone}

\begin{figure}[t]
\includegraphics[width=7cm,clip=true]{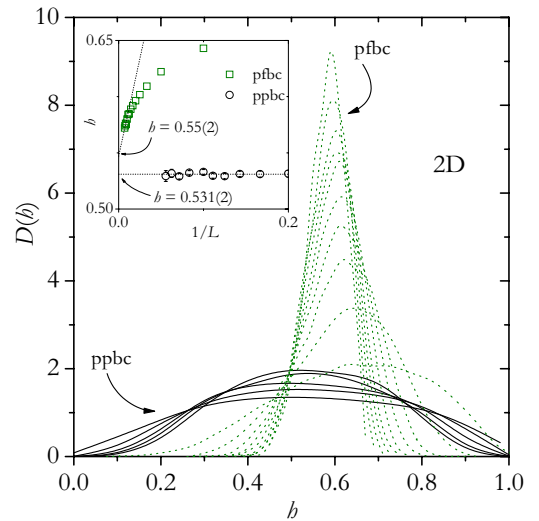}
\caption{(Color online) Distribution function $D(h)$ for the fraction of rigid bonds, for the 2D EAB model.
Curves for samples with ppbc of sizes $L=8$, $10$, $12$, $14$ and $16$,
and for samples with pfbc of sizes $L=10$, $20$, $30$, $40$, $50$,
$60$, $70$, $80$, $90$ and $100$ are shown (the curves are sharper for increasing $L$).
The inset shows the mean values of $h$ as a function of $1/L$ for both types
of boundary conditions.}\label{figure3}
\end{figure}

For the 2D EAB model with ppbc, the RL was calculated for $10^4$ samples for each size between $L=3$ and $L=12$,
$5 \times 10^3$ for $L=14$, $2 \times 10^3$ for $L=16$ and $10^3$ for $L=18$. We define the parameter $h$ as
the fraction of rigid bonds. Figure~\ref{figure3} shows the distribution function of $h$, $D(h)$,
for some of these sizes. Note that the curves get sharper as $L$ is increased: the standard deviation seems to decrease as $L^{-0.55(1)}$ and, as shown in the inset of Fig.~\ref{figure3}, the mean value converges quickly to $h = 0.531(2)$ (we have carried out a linear fit of points corresponding to the larger sizes).

\begin{figure}[t]
\includegraphics[width=7cm,clip=true]{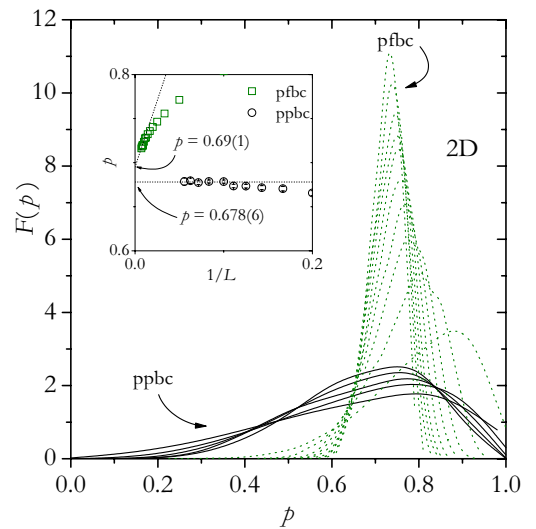}
\caption{(Color online) Distribution function $F(p)$ for the fraction of solidary spins, for the 2D EAB model.
The lattice sizes used are the same as in Fig.~\ref{figure3}.
The inset shows the mean values of $p$ as a function of $1/L$ for samples with
both types of boundary conditions.}\label{figure4}
\end{figure}

\begin{figure}[t]
\includegraphics[width=7cm,clip=true]{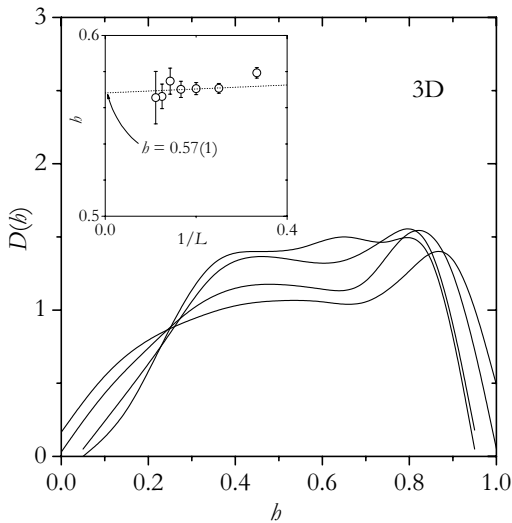}
\caption{Density distribution function $D(h)$ for the 3D EAB model.
Curves of sizes $L=4$, $6$, $8$ and $9$ are shown (the curves are narrower
for increasing $L$). The inset shows the mean values of $h$ as a function of $1/L$. }\label{figure5}
\end{figure}

\begin{figure}[t]
\includegraphics[width=7cm,clip=true]{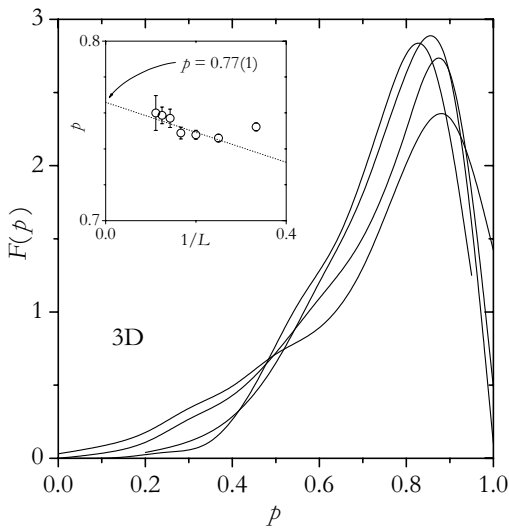}
\caption{Density distribution function $F(p)$ for the 3D EAB model.
The lattice sizes used are the same as in Fig.~\ref{figure5}.
The inset shows the mean values of $p$ as a function of $1/L$.}\label{figure6}
\end{figure}

Figure~\ref{figure3} also shows the distribution $D(h)$ for the 2D EAB
model with pfbc. In this case $10^4$ samples were calculated
for each size up to $L=50$, $7\times 10^3$ for $L=60$, $5\times 10^3$
for $L=70$, $3\times 10^3$ for $L=80$, $1.5\times 10^3$ for $L=90$,
$10^3$ for sizes between $L=100$ and $L=120$, and $5\times 10^2$ for $L=130$ and $L=140$
(the distributions shown in Fig.~\ref{figure3} are for sizes between
$L=10$ and $L=100$ only). In this case the standard deviation seems to decrease
as $L^{-0.56(1)}$ but, as the inset of Fig.~\ref{figure3} shows,
the mean value converges very slowly. This is probably due to finite size effects caused by the free boundary condition.  By fitting the points corresponding to the larger sizes, we conjecture that the mean value converges to $h = 0.55(2)$, which is above the value obtained for the ppbc (from now on, the error bars are omitted for clarity when they are equal or smaller than the symbol size). 

A similar behavior is observed in Fig.~\ref{figure4} for the distribution function $F(p)$, where $p$ is the fraction of S spins.  The mean values of this quantity for systems with both types of boundary conditions are shown in the inset. As before, using a linear fit of points corresponding to the larger sizes, we observe
a quick convergence to $p = 0.678(6)$ for samples with ppbc and a very slow convergence to a slightly larger value $p = 0.69(1)$ for samples with pfbc. Larger sample sizes with pfbc should be studied to determine more accurate values of fractions $h$ and $p$ in the thermodynamic limit.

The same analysis has been carried out for the 3D EAB model with ppbc. The RL was calculated for $10^4$ samples for sizes $L=3$ and $4$, $6\times 10^3$ for $L=5$, $3\times 10^3$ for $L=6$, $10^3$ for $L=7$ and $8$, and $2\times 10^2$ for $L=9$.
Figures~\ref{figure5} and \ref{figure6} show, respectively, the distributions $D(h)$ and $F(p)$ for sizes $L=4$, $6$, $8$ and $9$. As for 2D systems, in this case the curves seem to get sharper for larger $L$, and the standard deviation seems to decrease as $L^{-0.42(1)}$ for $D(h)$ and $L^{-0.68(2)}$ for $D(p)$. Although the lattice sizes are small, using a linear fit of points corresponding to the larger sizes, we can see in the insets that the mean values seem to converge to $h= 0.57(1)$ and $p = 0.77(1)$.

Previous numerical results show that the relative backbone size is slightly larger in 3D as compared to 2D. However, the most important difference between these two dimensionalities will arise in the internal structure of their respective backbones as we will show in the next subsection.  In particular, we do a detailed study of the percolation process exhibited by the backbone in the case of finite size lattices.

\subsection{\label{S3.3}Percolation of the backbone}

\begin{figure}[t]
\includegraphics[width=7cm,clip=true]{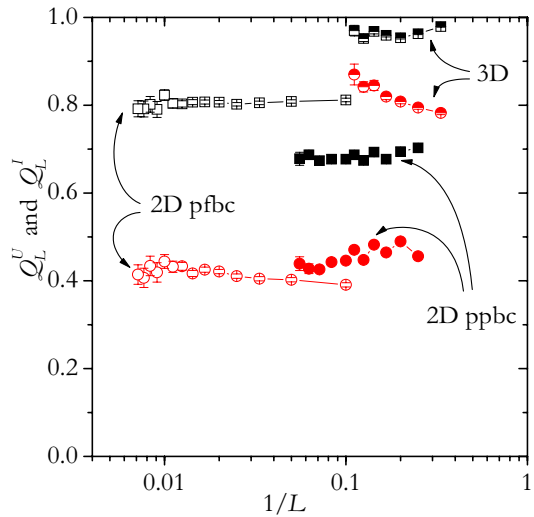}
\caption{(Color online) Percolation probabilities $Q^U_L$ (black squares) and $Q^I_L$ (red circles) as a
function of $1/L$.  Full and empty symbols correspond to, respectively,
RLs of 2D samples with ppbc and pfbc, while half-full symbols are
these percolation probabilities for the 3D system.}\label{figure7}
\end{figure}

A simple way of determining if the backbone percolates
is to calculate the percolation probability for different lattice sizes
and to extrapolate it to infinite $L$. Let us define $Q^U_L$ ($Q^I_L$) as the 
probability that the RL of the set of samples of size $L$ percolates along at least one lattice direction (percolates simultaneously along all independent lattice directions). \cite{Yonezawa1989} To conclude that the backbone percolates, these quantities should converge to $1$ in the thermodynamic limit.
Using the algorithm of Hoshen-Kopelman \cite{Hoshen1976} we have calculated $Q^U_L$ and
$Q^I_L$ for both 2D and 3D RLs. Figure~\ref{figure7} shows the percolation probabilities as function of $1/L$.
Whereas for 3D the results suggest that a percolation scenario can be likely, for 2D systems the curves do not show a clear tendency. Therefore, we have followed a different strategy to address this question in each case.

Following the line of reasoning given in Refs.~\onlinecite{Vogel1998a} and
\onlinecite{Vogel1998b}, we have studied the percolation probabilities as
functions of the variable $h$ (analog to the bond concentration in
random percolation).  For each linear lattice size $L$, we define $R^U_L(h)$ ($R^I_L(h)$) as the probability that the RLs having a fraction of rigid bonds between $h$ and $h + \Delta h$, percolates along at least one lattice direction (percolates simultaneously along all independent lattice directions).

\subsubsection{2D lattices}

\begin{figure}[t]
\includegraphics[width=7cm,clip=true]{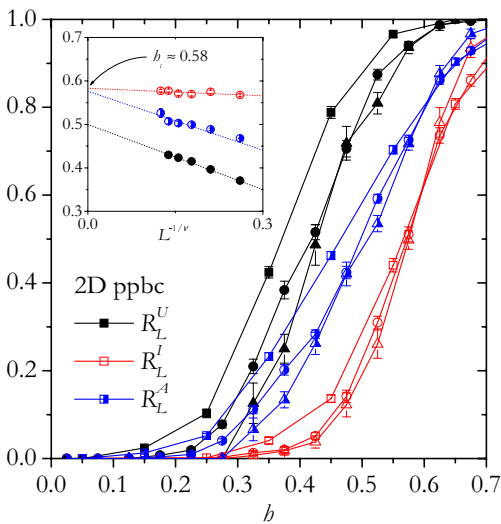}
\caption{(Color online) Percolation probabilities $R^U_L$ (black full symbols),
$R^I_L$ (red empty symbols) and $R^A_L$ (blue half-full symbols) as function of $h$, for
2D samples with ppbc.  Square, circle and triangle symbols are, respectively,
curves for lattice sizes $L=6$, $10$ and $16$.  The inset shows the corresponding
effective thresholds as a function of $L^{-1 /\nu}$ for sizes $L=6$, $8$, $10$,
$12$, $14$ and $16$.}\label{figure8}
\end{figure}

Figure~\ref{figure8} shows the functions $R^U_L(h)$, $R^I_L(h)$, and $R^A_L(h) \equiv [R^U_L(h)+R^I_L(h)]/2 $,
for three different lattice sizes of the 2D EAB model with ppbc. 
To obtain these curves, we have used a bin width of $\Delta h = 0.1$ up to $L=6$ and
$\Delta h = 0.05$ for larger sizes, and the value of each one of these probabilities
for samples with a fraction of rigid bonds between $h$ and $h + \Delta h$,
was assigned to the midpoint of the interval, i.e., to $h=h + \Delta h /2$.  Error bars were calculated using a bootstrap method. \cite{numerical}
The behavior of the curves in Fig.~\ref{figure8} suggests that the RL could be thought as the result of a bond-percolation process on a 2D lattice. \cite{Nota1}
The curves for $R^I_L(h)$ and $R^A_L(h)$ probabilities seem to cross at a concentration threshold $h_c$.
On the other hand, although the curves for $R^U_L(h)$ apparently cross,
this is hard to observe in Fig.~\ref{figure8} because of large finite-size effects.

To calculate a more precise value for the percolation threshold, we perform a
standard analysis of the data. \cite{Yonezawa1989}
First, each set of points is fitted with an error function using a least-mean-square method.
Then, the bond concentration at which the slope of the fitting curve is largest
is taken as an effective threshold $h^X_c (L)$, where $X$ denotes the percolation criterion used: $U$, $I$ or $A$. The $h^X_c (L)$ are expected to
follow the law \cite{Stauffer1985}
\begin{equation}
h^X_c(L)=h_c+C^X L^{-1 /\nu}.
\end{equation}
where $C^X$ is a non-universal constant and $\nu$ is the critical exponent associated to the correlation length.
The inset in Fig.~\ref{figure8} shows the effective thresholds as function of $L^{-1 /\nu}$,
where we have used the value of $\nu = 4/3$ of the 2D random percolation \cite{Stauffer1985}
(a justification of this choice is given below). To calculate an estimate of $h_c$,
we extrapolate towards the thermodynamic limit by means of a linear fit.
For $h^I_c(L)$ and $h^A_c(L)$ we obtain the limits $0.58(1)$ and $0.58(3)$ respectively.  As before, these quantities do not agree with the extrapolated value for $h^U_c(L)$, $0.49(2)$, probably because the percolation criterion $U$ is more sensitive to finite size effects.

\begin{figure}[t]
\includegraphics[width=7cm,clip=true]{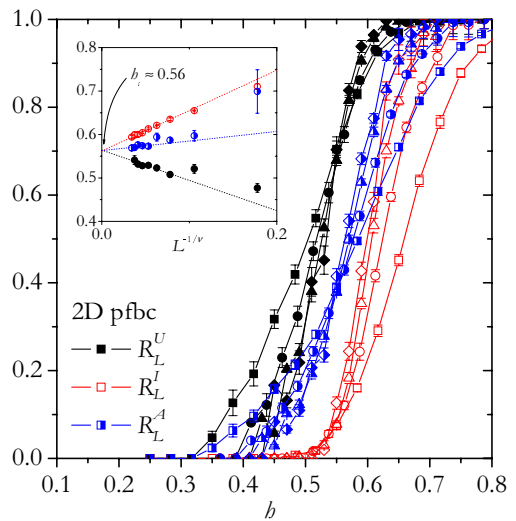}
\caption{(Color online) Percolation probabilities $R^U_L$ (black full symbols),
$R^I_L$ (red empty symbols) and $R^A_L$ (blue half-full symbols) as a function of $h$, for
2D samples with pfbc.  Square, circle, triangle and rhombus symbols are, respectively,
curves for lattice sizes $L=20$, $40$, $60$ and $80$.  The inset shows the corresponding
effective thresholds as a function of $L^{-1 /\nu}$ for lattices sizes up to $L=90$.}\label{figure9}
\end{figure}

Thus, the results obtained for small samples with ppbc suggest the existence of a
critical concentration $h_c \approx 0.58$ in 2D. Note that this value is different from the concentration threshold of the random-bond percolation in
the square lattice, $\rho_c = 0.5$, \cite{Stauffer1985} which was to be expected because the bonds of the RL are not independently and randomly placed on the lattice. Another important difference is that in the rigid bond percolation process the concentration of rigid bonds cannot be freely varied for very large system sizes, because the distribution $D(h)$ tends to a delta function. The simulations for the 2D samples with ppbc described above seem to indicate that this delta function is placed at $h=0.531(2)$, which would imply that the RL does not percolate in the thermodynamic limit. However, the closeness of the values obtained and the small sizes considered do not allow us to discard the possibility that the delta function is in fact placed at $h_c$.

Eventually, the situation should be clearer when considering 2D samples with pfbc, because larger system sizes are available. Unfortunately we show below that the analysis leads to a similar conclusion. Figure~\ref{figure9} shows the functions $R^U_L(h)$, $R^I_L(h)$ and $R^A_L(h)$
for four different sizes. We have used a bin width ranged from $\Delta h = 1/30$
(samples with $L=20$) to $\Delta h = 1/50$ (samples with $L \ge 60$).
The differences between Figs.~\ref{figure8} and \ref{figure9} are evident.
In this case the crossing of the $R^I_L(h)$ curves happens at a very
low value of probability. This is not surprising since samples with pfbc
are very anisotropic: for a given sample, the percolation probability in the
lattice direction $x$ (where periodic boundary conditions are used)
is larger than in the lattice direction $y$ (where free boundary conditions
are imposed). In fact, $R^x_L(h) \approx R^U_L(h)$ and $R^y_L(h) \approx R^I_L(h)$.
Inset in Fig.~\ref{figure9} shows the effective thresholds as function of $L^{-1 /\nu}$
(again we have used $\nu = 4/3$).  By extrapolating toward the thermodynamic limit
we obtain the limits $0.56(1)$, $0.562(5)$ and $0.56(2)$ for, respectively,
$h^U_c(L)$, $h^I_c(L)$ and $h^A_c(L)$.  This critical concentration of
$h_c \approx 0.56$ is very close to the mean value of $h=0.55(2)$ for
2D samples with pfbc.  Then, a similar scenario to the one previously found for
the samples with ppbc is obtained: even though $h_c$ is slightly above the asymptotic mean value of $h$, they are so close that they fall within the error bar of each other. Therefore, from this study no definite conclusion can be drawn for the 2D cases, and the evidence seems to indicate that the sample size needed for achieving a definite answer on the percolation of the backbone is orders of magnitude larger than the ones available to us. Note that for the 2D EAB model finite size effects also affect the determination of many other quantities. For instance, to be reasonably sure that the stiffness exponent of the defect energy vanishes, samples of up to $L=480$ had to be analyzed. \cite{Hartmann2001}

\begin{figure}[t]
\includegraphics[width=7cm,clip=true]{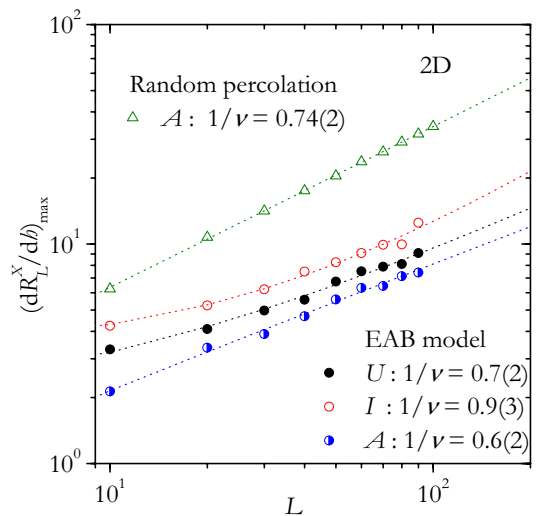}
\caption{(Color online) $\left( \frac{dR^X_L}{dh} \right)_{\mathrm{max}}$
as a function of $L$ for the 2D EAB model with pfbc (circles) and for the random-bond
percolation model in the square lattice (triangles). The different percolation criteria
are indicated in the figure. The dotted lines are best fits obtained using Eq.~(\ref{scaling}).} \label{figure10}
\end{figure}

Now, we deal with the problem of determining the universality class of the
percolation process and the main characteristics of the RLs internal structure.
Because it is necessary to analyze large samples, in most of the cases we have
restricted the study to samples with pfbc. In the fits described above we have supposed that $\nu=4/3$, i.e. that the universality belongs to the 2D random
percolation. A first indication that this is in fact the case,
is that the linear fits in inset of Fig.~\ref{figure9} intersect very close to the ordinate axis, for $\nu = 4/3$.
But $\nu$ can be evaluated using the following
expression: \cite{Stauffer1985}
\begin{equation}
\left( \frac{dR^X_L}{dh} \right)_{\mathrm{max}} \propto L^{1 /\nu}.
\end{equation}
In Fig.~\ref{figure10} we show the maximum of this derivative for
the three types of percolation probability as functions of $L$.  Only samples
with pfbc were used. Due to the finite-size effects displayed in this figure, we have chosen
to fit the data using a simple scaling function with an additional correction term,
\begin{equation}
f(L)=a+b L^{1 /\nu},  \label{scaling}
\end{equation}
where $a$ and $b$ are constants.
The values of $1/\nu$ obtained for the percolation criterion $U$, $I$
and $A$ are, respectively, $0.7(2)$, $0.9(3)$ and $0.6(2)$.  Although
the error bars are large, these values are compatible with the
2D random percolation universality class.  For comparison, we show
in Fig.~\ref{figure10} the results obtained for the random-bond percolation in
the square lattice (only criterion $A$ is shown).  Using the same scaling
function and samples up to $L=100$ we obtain $1/\nu=0.74(2)$. \cite{Nota3}

\begin{figure}[t]
\includegraphics[width=7cm,clip=true]{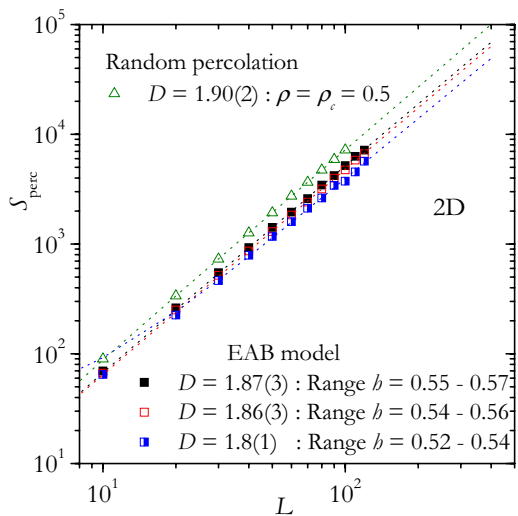}
\caption{(Color online) Mean size of the percolating cluster as a function of $L$
for the 2D EAB model with pfbc (squares) and for the random-bond
percolation model in the square lattice (triangles). The used ranges of $h$
are indicated in the figure. The dotted lines are best fits obtained using Eq.~(\ref{scaling}).} \label{figure11}
\end{figure}

To characterize the topology of the backbone, we turn now to $D$, the fractal dimension of the percolating cluster.
It is defined by
\begin{equation}
S_{\mathrm{perc}} \propto L^D , \label{Sper}
\end{equation}
where $S_{\mathrm{perc}}$ represents the mean number of elements (in this case rigid bonds)
which form the spanning cluster. We have calculated $S_{\mathrm{perc}}$ for the samples that have a percolating cluster, for three different ranges of $h$: one centered in
$h=0.56$, the percolation threshold calculated for the EAB model with pfbc, and other two
ranges centered in $h=0.55$ and $h=0.53$, the mean values of $h$ calculated for samples
with pfbc and ppbc, respectively.  We plot in Fig.~\ref{figure11} $S_{\mathrm{perc}}$ as a
function of $L$ for these ranges and again, for comparison, the results obtained for
the random-bond percolation in the square lattice at $\rho_c=0.5$.  In all the cases we fitted the data
with the scaling function, Eq.~(\ref{scaling}).  The values obtained for the fractal dimension are shown
in Fig.~\ref{figure11}. It can be noticed that all these values are compatible with $D=91/48 \approx 1.896$,
the fractal dimension of the percolating cluster in the 2D random percolation universality class. \cite{Stauffer1985}

\begin{figure}[t]
\includegraphics[width=7cm,clip=true]{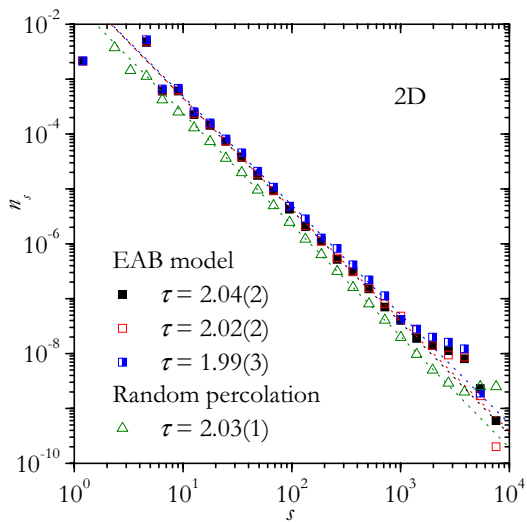}
\caption{(Color online) Cluster number distribution for 2D samples of the EAB model with pfbc and
for the random-bond percolation in the square lattice.  In both cases the lattice size is $L=120$.
Symbols are the same as in Fig.~\ref{figure11}. } \label{figure12}
\end{figure}

In Fig.~\ref{figure12} we show the cluster number distribution (i.e. the number of
clusters of size $s$), $n_s$, for the same three ranges of $h$ previously studied and for
the 2D random bond percolation in the square lattice at $\rho_c=0.5$.  At the critical bond concentration
it is expected that this distribution follows a power law
\begin{equation}
n_s \propto s^{-\tau},
\end{equation}
where $\tau$ is a critical exponent.  By fitting all the curves in Fig.~\ref{figure12},
we obtain different values of $\tau$ which are very close to $\tau=187/91 \approx 2.05$,
the corresponding exponent for the 2D random percolation universality class. \cite{Stauffer1985}

\subsubsection{3D lattices}

\begin{figure}[t]
\includegraphics[width=7cm,clip=true]{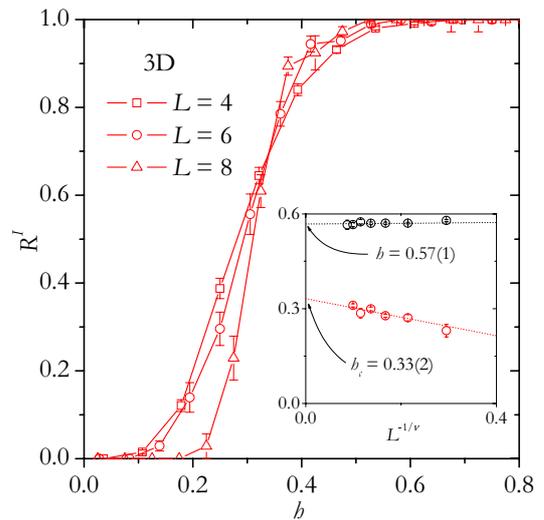}
\caption{(Color online) Percolation probability $R^I_L$ versus $h$, for
the 3D EAB model.  Square, circle and triangle symbols are, respectively,
curves for lattice sizes $L=4$, $6$ and $8$. The inset shows the corresponding
effective threshold and the mean value of $h$ as a function of $L^{-1 /\nu}$, where
we have used $\nu = 0.9$.} \label{figure13}
\end{figure}

Even though Fig.~\ref{figure7} suggests that for 3D systems the backbone percolates, we have carried out the same analysis as for 2D systems to confirm this observation. Figure~\ref{figure13} shows the curves of $R^I_L$ versus $h$ for three lattices sizes. The other percolation criteria have not been included because, as Fig.~\ref{figure8} shows for 2D systems with ppbc, they are much more sensitive to finite size effects which are rather large for the small systems analyzed.
The inset in Fig.~\ref{figure13} shows the effective thresholds $h^I_c( L)$ as function of $L^{-1 /\nu}$ for all available system sizes. Similarly to the 2D case we have chosen $\nu = 0.9$, the value of 3D random percolation (this choice is justified below). \cite{Stauffer1985} In the thermodynamic
limit we obtain a critical threshold of $h_c=0.33(2)$, which is much smaller than the mean
fraction of rigid bonds, $h=0.57(1)$ (see inset in Fig.~\ref{figure13} for a comparison).
In addition, note that as in 2D, the RL percolates at a larger concentration value than
the corresponding one for the random-bond percolation in the simple cubic
lattice, $\rho_c \approx 0.2488$. \cite{Lorenz1998}

\begin{figure}[t]
\includegraphics[width=7cm,clip=true]{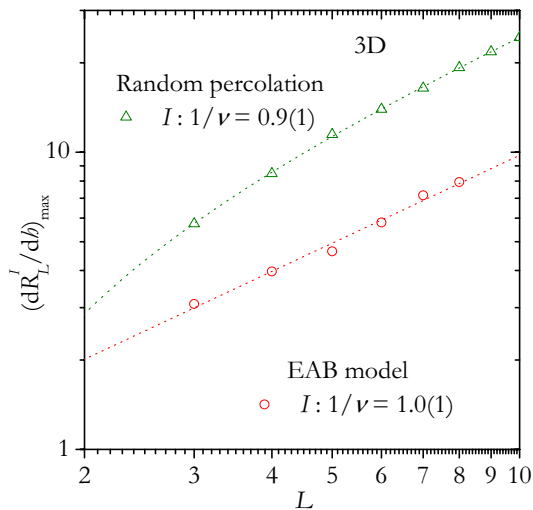}
\caption{(Color online) $\left( \frac{dR^I_L}{dh} \right)_{\mathrm{max}}$
as a function of $L$ for the 3D EAB model (circles) and for the random-bond
percolation model in the simple cubic lattice (triangles). } \label{figure14}
\end{figure}

To determine the universality class of the percolation process we proceed as for the 2D case. Figure~\ref{figure14} shows the data for $\left( \frac{dR^X_L}{dh} \right)_{\mathrm{max}}$, fitted by the function given in Eq.~(\ref{scaling}). The value obtained for the exponent is $1/\nu=1.0(1)$, which is compatible with the value for random percolation. The large error in the value of the exponent is justified by the small sizes used. Note that errors of the same amount are obtained in the case of random percolation in 3D if the exponent is calculated using the same criterion for systems of the same sizes (see Fig.~\ref{figure14}).

\begin{figure}[t]
\includegraphics[width=7cm,clip=true]{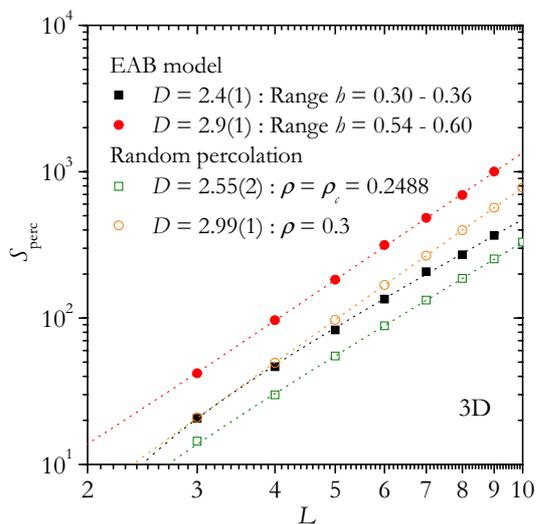}
\caption{(Color online) Mean size of the percolating cluster as a function of $L$
for the 3D EAB model (full symbols) and for the random-bond
percolation model in the simple cubic lattice (open symbols). The used ranges or
concentrations are indicated in the figure.} \label{figure15}
\end{figure}

Figure~\ref{figure15} shows the data for the size of the percolating cluster as a function of $L$, which allows us to obtain the fractal dimension of the backbone, using Eq.~(\ref{Sper}). Data are shown for two ranges of $h$: around the mean value of $h$ and around the critical value obtained in Fig.~\ref{figure13}. In this last case the value obtained for the fractal dimension is $D=2.4(1)$, which is very close to $D=2.53$, the corresponding value for 3D random percolation at the percolation threshold. \cite{Stauffer1985} For the range of $h$ centered at the mean value, the fractal dimension obtained is also very close to $D=3$, the random percolation value for a bond concentration $\rho>\rho_c$. To gauge the influence of finite size effects, we have included in the figure the data obtained for random bond percolation in the cubic lattice for the same system sizes, for concentrations $\rho = \rho_c \approx 0.2488$ and $\rho=0.3$.

\begin{figure}[t]
\includegraphics[width=7cm,clip=true]{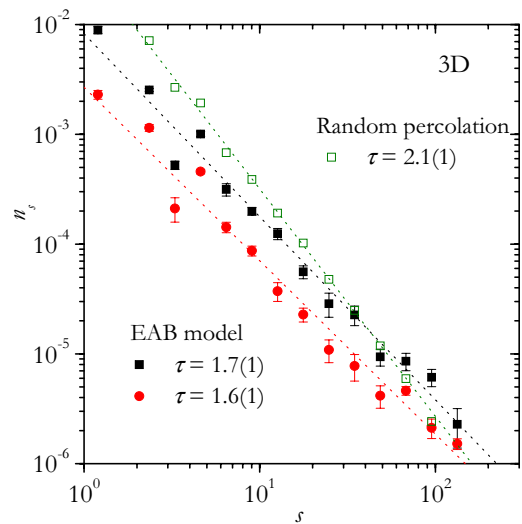}
\caption{(Color online) Cluster number distribution for 3D samples of the EAB model with $L=8$ and
for the random-bond percolation in the simple cubic lattice with $L=8$.
Symbols are the same as in Fig.~\ref{figure15}. } \label{figure16}
\end{figure}

Even though the evidence shown above seems to indicate that the RL
is in the same universality class as random percolation, the data
for the distribution of cluster sizes, shown in
Fig.~\ref{figure16}, do not fit into this picture. At the
percolation threshold calculated above for the RL, we obtain
$\tau=1.7(1)$ which is clearly different from the accepted value,
$\tau = 2.2$ (Ref. \onlinecite{Stauffer1985}) and even from the value obtained
when calculated over the same small sizes used for the RL.
Furthermore, for $h$ close to the asymptotic mean value, which is
well above the percolation threshold, we find that the data for
$n_s$ still seem to follow a power law, with a similar exponent
$\tau =1.6(1)$. It is well known, however, that this is not the
case for random percolation above the threshold, not even when
concentration is very close to it.

\subsection{\label{S3.4}Average energy}

\begin{figure}[t]
\includegraphics[width=7cm,clip=true]{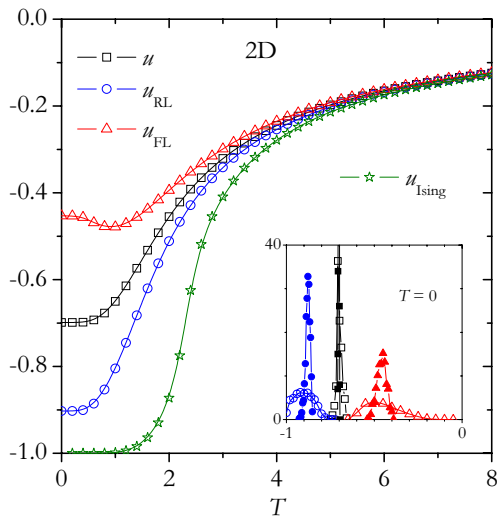}
\caption{(Color online) Energies per bond for the whole system $u(T)$, the RL $u_\mathrm{RL}(T)$ and the FL $u_\mathrm{FL}(T)$, for the 2D EAB model with ppbc and $L=16$. $u_\mathrm{Ising}$ is the energy per bond for a ferromagnetic Ising model with ppbc and the same size. The inset shows the histograms of these three energies for $T=0$. Empty symbols in inset are the same as in the main figure. Full symbols correspond to the same quantities calculated for the 2D EAB model with ppbc and $L=100$.}\label{figure17}
\end{figure}

In this section we study the contribution of the backbone to the energy of the system. Let us define $u(T)$,
$u_\mathrm{RL}(T)$ and $u_\mathrm{FL}(T)$ as the average energies per bond at temperature $T$ of, respectively, the whole system, the RL and the FL. $u_\mathrm{RL}(T)$ and $u_\mathrm{FL}(T)$ are calculated by restricting the Edwards-Anderson Hamiltonian to the bonds and spins that belong to each region, and dividing by the corresponding number of bonds. For the 2D EAB model with ppbc, we have calculated these energies using a parallel tempering algorithm, \cite{Geyer1991,Hukushima1996} with $m=40$ replicas of the system for temperatures decreasing from $T=8.0$ all the way to $T=0.2$. $2000$ samples of size $L=16$ were equilibrated using $10^6$ parallel tempering steps (PTS), where a PTS consists of $m\times N$ elementary steps of standard Monte Carlo (Metropolis) and only one replica exchange. The energy averages were performed using the same number of PTS. The resulting energies are shown in Fig.~\ref{figure17}. The values at $T=0$ are the GS energies calculated as in the previous section using parallel tempering (but without equilibrating to make the algorithm faster). For comparison, the figure also shows the energy per bond for the ferromagnetic Ising model, with $L=16$. As 2D EAB systems with pfbc are much harder to equilibrate, we have only calculated the energies per bond for the case of $T=0$, using a Blossom algorithm.

The most interesting feature is that, whereas for high temperatures all the energies coincide, when the temperature is progressively lowered the curves separate, with $u_\mathrm{FL}(T) > u(T) > u_\mathrm{RL}(T)$.  We have checked that this happens not only on the average, but it is also true for single samples, as the histograms of these three energies for $T=0$ shown in the inset of Fig.~\ref{figure17} confirm. This is an indication that frustration is not distributed homogeneously in the system; instead, it is concentrated mainly in the FL. Note also that the decrease in $u_\mathrm{FL}(T)$ is not monotonous but it displays a minimum at finite temperature. On the other hand, the energy of the RL does decrease monotonously to its minimum at $T=0$, $u_\mathrm{RL}(0) \approx 0.9$ (indicating that the fraction of frustrated bonds inside the RL is of only $\approx 5 \% $). Interestingly, this value can be considered as the GS energy for the rigid lattice: it is the value that is obtained when all the bonds and spins that do not belong to the RL are eliminated and the GS of the resulting system is calculated. We have checked that this happens for all the samples analyzed. If, on the other hand, the same is done for the FL, we have checked that $u_\mathrm{FL}(0)$ is always much larger than the GS energy of the new system where bonds and spins that do not belong to the FL are eliminated. It can therefore be said that the FL is in an {\em excited} phase.\cite{Roma2007a}

\begin{figure}[t]
\includegraphics[width=7cm,clip=true]{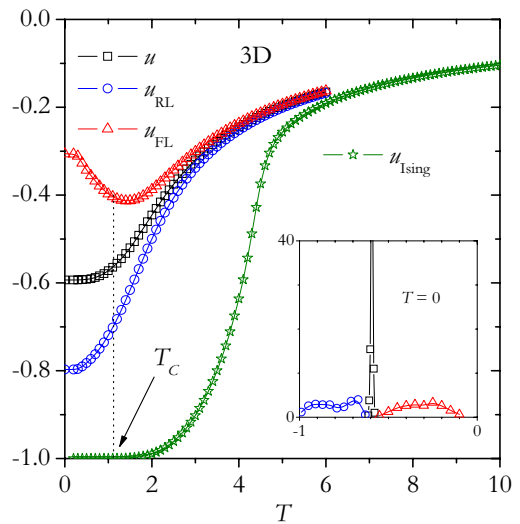}
\caption{(Color online) Energies per bond for the whole system $u(T)$, the RL $u_\mathrm{RL}(T)$ and the FL $u_\mathrm{FL}(T)$, for the 3D EAB model with $L=8$. $u_\mathrm{Ising}$ is the energy per bond for a ferromagnetic Ising model with ppbc and the same size. The inset shows the histograms of these three energies for $T=0$. Symbols in the inset are the same as in the main figure.}\label{figure18}
\end{figure}

We have carried out similar calculations for the 3D EAB model with $L=8$, using $1000$ samples which were equilibrated using $10^6$ PTS with $m=60$ replicas, for temperatures between $T=6$ and $T=0.2$. Figure~\ref{figure18} shows that the main features of the curves of the three energies are the same as for the 2D case. Note also that the separation of the curves happens at temperatures well above the critical one $T_{\mathrm{c}} \approx 1.12$. \cite{Katzgraber2006} $u_\mathrm{RL}(0)$ decreases to a value of $u_\mathrm{RL}(0) \approx -0.8$, indicating that the fraction of frustrated bonds inside the RL is of only $\approx 10 \% $.

\begin{figure}[t]
\includegraphics[width=\linewidth,clip=true]{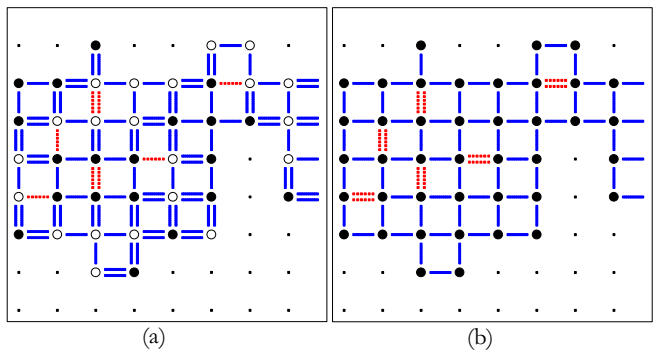}
\caption{(Color online) Gauge transformation. Panel (a) shows the largest cluster of the RL of the 2D sample in Fig.~\ref{figure1}, whereas panel (b) shows the result of applying the gauge transformation to this structure. Symbols are the same as in Fig.~\ref{figure1}.}\label{figure19}
\end{figure}

The fact that the RL displays very little frustration suggests that its behavior can be associated to certain extent 
to a ferromagnetic phase. To see this, it is useful to consider what happens to the RL when a gauge transformation is applied. This transformation, which leaves the Hamiltonian invariant, consists in flipping one spin, as well as flipping the sign of the surrounding bonds, and repeating these steps until all spins have the same direction.\cite{Vannimenus1977} Figure~\ref{figure19}(a) shows the largest cluster of the RL of the same 2D sample as in Fig.~\ref{figure1}, and Fig.~\ref{figure19}(b) shows the result of the gauge transformation. Note that the frustrated bonds are conserved, but all of them were mapped to antiferromagnetic bonds. Satisfied bonds on the other hand are mapped to ferromagnetic bonds. Thus, using the results given above, the transformed structures have an average concentration of antiferromagnetic bonds of $x=0.05$ and $x=0.1$, respectively, for the 2D and 3D EAB model. It is instructive to compare the transformed backbone with the Edwards Anderson model for a bimodal distribution $J_{i,j} = x \delta (J_{i,j}+1) + (1-x) \delta (J_{i,j}-1)$, (also known as random bond Ising model). This model has a ferromagnetic phase for low concentrations up to $x_c = 0.104(1)$ and $x_c = 0.222(5)$ in 2D and 3D, respectively . \cite{Kawashima1997,Hartmann1999a}  Note that these concentrations are larger (almost by a factor of $2$) than the ones we have found for the RL both in 2D and 3D. This suggests that the backbone could be considered as a ferromagnetic component embedded in the spin glass. This analogy, however, is not complete because in the transformed structure the antiferromagnetic bonds cannot be necessarily considered as random, since
the frustrated bonds of the RL are correlated. Furthermore, for this analogy to be meaningful, the RL should be a percolating structure. Thus, the analogy is better for 3D than for 2D systems, where we have shown that there is no solid evidence of percolation.

%.....................................................................
\section{\label{Dis}DISCUSSION}

In this section we use the results obtained in this and previous works, to try to understand the role that the backbone plays in the behavior of some spin glass systems, and in general in disordered and frustrated systems.

Because of the fact that the determination of the GS of the EAB model is an NP-hard problem, in this work all numerical studies were restricted to relatively small samples. Nevertheless, the results obtained suggest that whereas in 2D there is no evidence that the RL percolates, in 3D percolation of the RL seems to be the most probable scenario. Taking into account the fact that in 2D the critical temperature is vanishing whereas for 3D it is known that $T_c>0$, it can be conjectured that the properties of the low temperature phase must be strongly correlated with the properties of the backbone (characterized by the RL, in this work).

The spatial distribution of frustration found in EAB systems is also very interesting. We have observed that frustration is not homogeneously distributed in the system: it is mostly concentrated in the FL, whose energy per bond $u_\mathrm{FL}(T)$ is thus much larger than the corresponding energy of the RL, $u_\mathrm{RL}(T)$. Moreover, Figs.~\ref{figure17} and \ref{figure18} show that $u_\mathrm{FL}(T)$ has a minimum value for nonvanishing $T$ and that in 3D at $T=0$ it takes a value that is close to the energies of the whole system in the paramagnetic phase (in the 2D case the value coincides with the energy of the system for $T \approx 1.3$). This suggests that the FL can be thought as a subsystem in an excited state. \cite{Roma2007a} This impression is supported by the fact that, when considered as an isolated system, the GS energy of the FL is smaller than $u_\mathrm{FL}(T)$. Furthermore, this also suggests that the behavior of the FL is closely related to that of a paramagnet, even at vanishing temperatures.

When compared with the ferromagnetic phase of the random bond Ising model at low concentrations of antiferromagnetic bonds, the low frustration that we find in the RL seems to indicate that the RL could share many of the properties of a ferromagnet (even though in 3D there is no spontaneous magnetization of the backbone, for $T<T_c$, the clusters of S spins align in one of two possible directions). It must be said, however, that this is only a conjecture, because, as mentioned in the previous section, one cannot be sure that the gauge transformation maps a typical distribution of frustrated bonds to a typical distribution of random bonds in the random bond Ising model, because of the correlations that arise between frustrated bonds of the RL.

Recently, the fact that the RL and the FL have very different properties has also been verified in other contexts. \cite{Roma2006,Roma2007b,Roma2010} For instance, the size dependence of the defect energy in 3D shows that the RL has a very large stability (stiffness exponent $\theta_\mathrm{RL} = 2.59(2)$), comparable with that of a ferromagnet (stiffness exponent $\theta = 2$ for the Ising model in 3D). This large stability must also be compared to the stability of the system as a whole, which is much lower ($\theta \approx 0.2$). \cite{Hartmann1999b} This arises as a compensation between the large stability of the RL and the instability of the FL. The relevance of the separation of the system in RL and FL can also be appreciated in the out of equilibrium dynamics. It has been observed that the distribution of spin mean flipping times has two peaks, corresponding to groups of fast and slow spins.\cite{Ricci2000}  But a detailed analysis of the contributions of the RL and FL reveals that most of the slow spins belong to the RL and most of the fast spins belong to the FL. \cite{Roma2006,Roma2010}

Further evidence of the different roles that the RL and FL play in the physics of the system is obtained in the study of the violation of the FDT.\cite{Roma2007b} Whereas inside the RL a coarsening-like behavior is observed, the FL asymptotically follows the FDT (a behavior characteristic of a paramagnetic phase). All these results support the idea that in the EAB model the system can be divided into two regions that have very different properties: one behaves mostly like a ferromagnet (RL) whereas the other behaves more like a paramagnet (FL).

Evidently, the characterization of the backbone given in this and other works can only be applied to systems having a degenerate GS. But, in general, systems with a continuous distribution of bonds have at most one GS (or two if we allow a global spin flip), where the RL would encompass the whole system. To extend the notion of backbone to such systems a more general definition of rigidity should be given. In particular the `rigidity' of each bond should be associated to a parameter taking a continuum of values, instead of only two (rigid-flexible) as in the EAB.

One possible generalization is as follows. For a bond $J_{ij}$ we define its rigidity $r_{ij}=U^*_{ij}-U$, where $U$ is the GS energy of the sample and $U^*_{ij}$ is the lowest energy for which the condition of the bond $J_{ij}$ is frustrated (satisfied) if it is satisfied (frustrated) in the GS. As shown below, this seems to be a very reasonable generalization. The algorithm to find the rigidity of each bond is very similar to the RLSA: after finding the GS of the system, one of the spins $i$ and $j$ is flipped, and then both are `frozen' and the lowest energy of the constrained system is calculated. This gives $U^*_{ij}$, the energy of the lowest excited state where the bond $J_{ij}$ is in a different condition than in the GS.

\begin{figure}[t]
\includegraphics[width=7cm,clip=true]{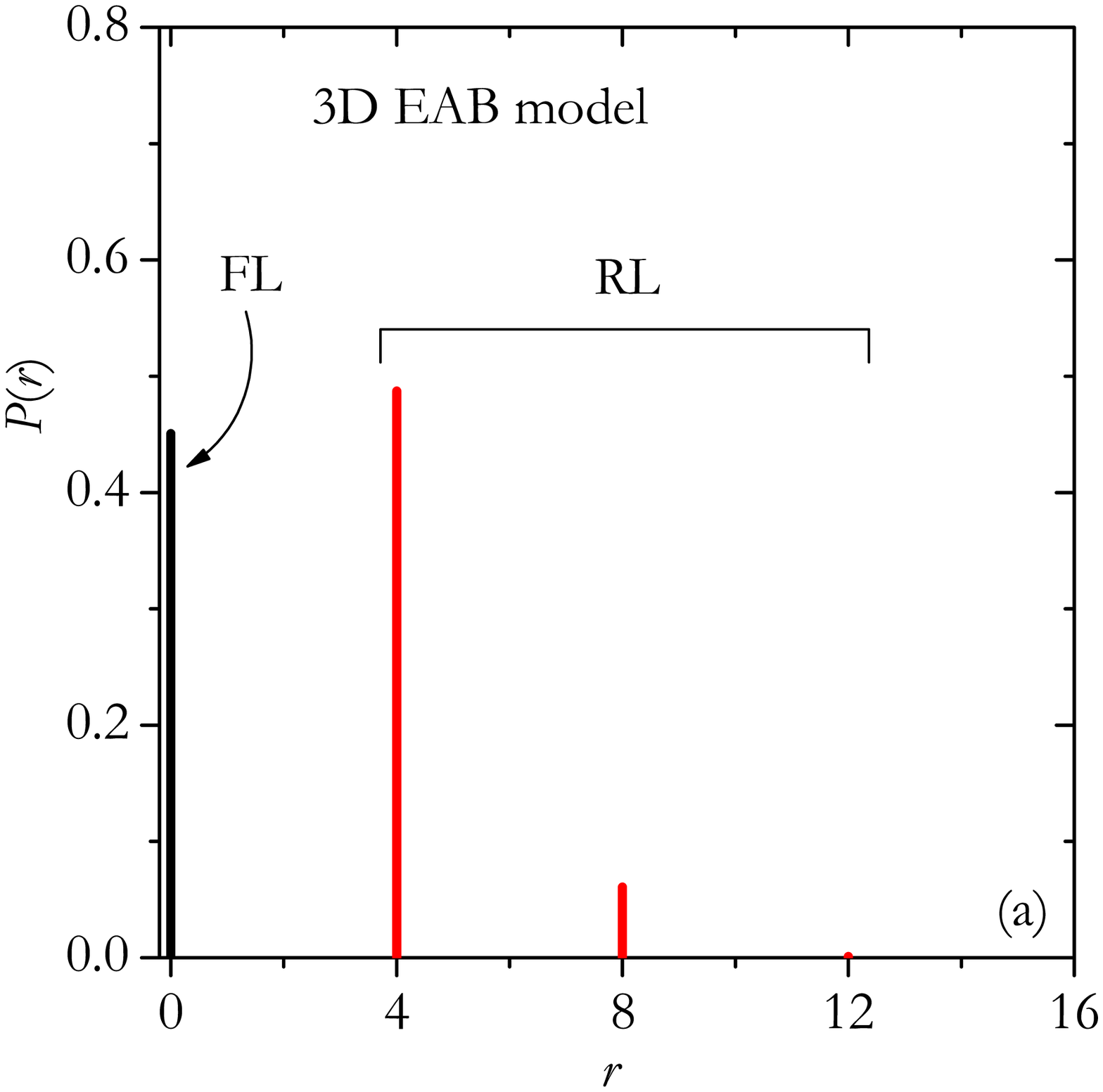}
\includegraphics[width=7cm,clip=true]{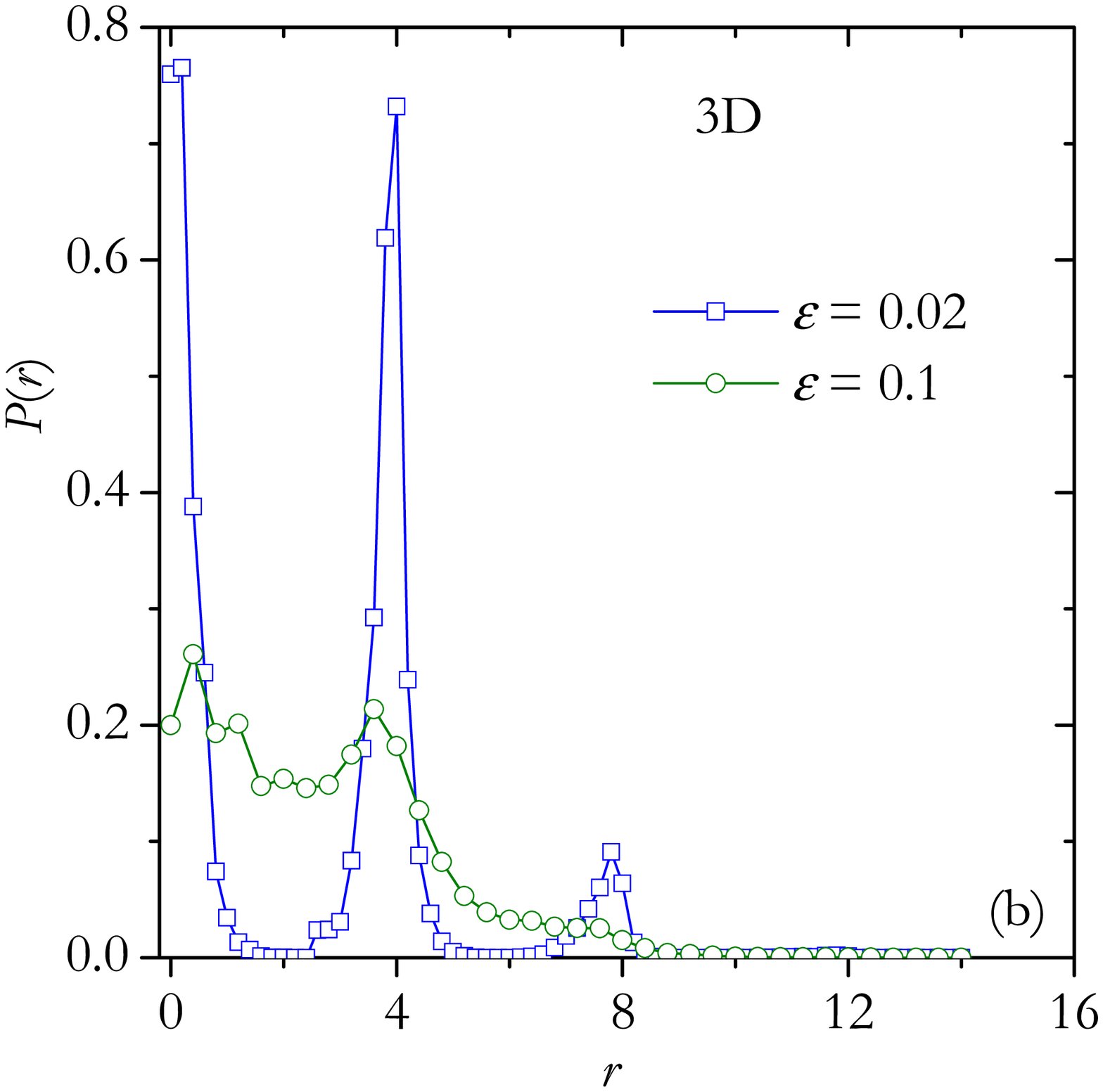}
\caption{(Color online) Rigidity distribution $P(r)$ for (a) the 3D EAB model and (b) the 3D EAB-$\epsilon$ model with $\epsilon=0.02$ and $0.1$. In both cases we have analyzed $10^3$ samples of systems with $L=6$.} \label{figure20}
\end{figure}

We have performed a preliminary study of the distribution of rigidities for different bond distributions in the Edwards-Anderson model (a more complete analysis will be presented elsewhere). First, note that in the 3D EAB model the rigidity $r$ can only take 4 values: 0 (bonds that belong to the FL), 4, 8 or 12 (bonds belonging to the RL). In 2D bonds in the RL can only have $r=4$ or $r=8$.  Figure~\ref{figure20} (a) shows the distribution of these values for 3D samples with $L=6$, for which the average rigidity is $\bar{r} \approx 2.45$.

Figure~\ref{figure20} (a) shows that the concept of rigidity allows us to detect the heterogeneity inside of the RL. Nevertheless, it is interesting to notice that also the FL is not a homogeneous lattice. In fact, a flexible bond which changes its condition (satisfied or frustrated) in only a few GS configurations is expected to have a more rigid behavior than flexible bonds that changed in many GS configurations.  A study of the heterogeneous character of the FL could be carried out by uniform sampling of the GS using, for example, the algorithm proposed in Ref.~\onlinecite{Thomas2009}.

To understand what changes when continuous distributions of bonds are considered, it is useful to choose a distribution that has the EAB as a limit case. One obvious choice is a distribution that consists on the superposition of two Gaussian function of width (variance) $\epsilon$ centered at $J=1$ and $J=-1$. We call this the EAB-$\epsilon$ model. For $0<\epsilon \ll 1$ it is reasonable to expect that, even though it has only one GS configuration, the properties of the system will not be very different from the corresponding ones for the EAB model. Figure~\ref{figure20} (b) shows that for $\epsilon=0.02$ the rigidity distribution is very similar to the one obtained for the EAB, with four peaks centered at $r=0$, $4$, $8$ and $12$. For this value of $\epsilon$ the sharpness of the peaks still allows us to divide the system into four well defined components. This, in turn, makes it possible to define a backbone for this continuous distribution of bonds. When $\epsilon$ is increased, the peaks become necessarily less sharp (see Fig.~\ref{figure20} (b)). The average rigidity, however, keeps almost constant: we have $\bar{r} \approx 2.47$ for $\epsilon=0.02$ and $\bar{r} \approx 2.66$ for $\epsilon=0.1$.

\begin{figure}[t]
\includegraphics[width=7cm,clip=true]{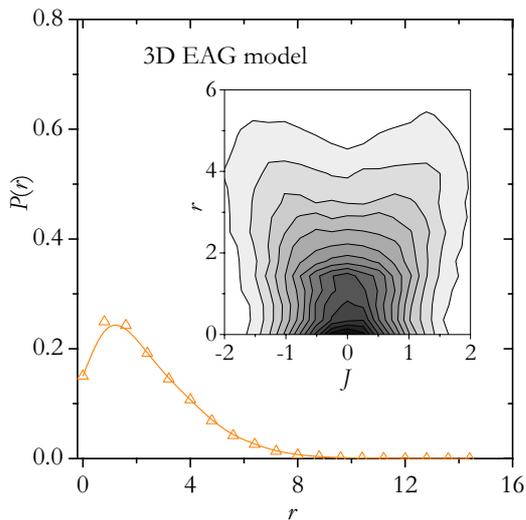}
\caption{(Color online) Rigidity distribution $P(r)$ for the 3D EAG model calculated for $10^3$ samples of systems with $L=6$. The inset shows a map plot of the probability that a bond of strength $J$ has a rigidity $r$. Darker levels of gray indicate larger probabilities.
} \label{figure21}
\end{figure}

It is interesting to see what happens when the rigidity of the EAG model is analyzed. Figure~\ref{figure21} shows that $P(r)$ is not very different from the rigidity distributions displayed in Fig.~\ref{figure20} (b) for $\epsilon=0.1$. Here the average rigidity is $\bar{r} \approx 2.31$. The inset of Fig.~\ref{figure21} is a map plot of the probability that a bond with value $J$ has rigidity $r$. The rectangular shape of the map plot shows that that the strength of the bond is only very weakly related to its rigidity. For example, bonds having $|J|<1.5$ have the same probability of having $r=4$. Note that the same situation arises in the EAB where the RL and the FL have identical proportions of ferromagnetic and antiferromagnetic bonds. Thus, we see that neither the strength nor the sign of a bond alone can account for its rigidity.

On the other hand, we argue that the rigidity of bond $J_{ij}$ is a quantity that can give an idea of the `effective interaction' between spins $i$ and $j$, in systems with quenched disorder, as the 3D EAB. This conjecture is supported by the fact that in the out-of-equilibrium dynamics of this system it is observed  \cite{Roma2006,Roma2010} that non solidary spins flip as if they were embedded in a paramagnetic phase, which indicates that the effective interaction between them is very small. For solidary spins, on the other hand, the effective interaction seems to be much stronger because they flip much less often.

Even though the numerical results described in this paper were obtained for the EAB model, the generalization advanced in the previous paragraphs seems to indicate that the concept of backbone can be used in other disordered and frustrated systems. This structure could be obtained by thoroughly studying and comparing the configurations of the fundamental state as well as of the first excited levels.

As mentioned before, we think that the idea of separately studying the contribution of the backbone to the different physical quantities of a system could be very useful for a better understanding of the low temperature phase of spin glasses. In particular, it could shine some light on the long standing controversy between the droplet picture and the RSB picture. The droplet picture postulates that below the critical temperature a spin glass is essentially like a ferromagnet, in the sense that it should have a trivial energy landscape. If this picture was correct, it could be thought that what is happening is that the backbone (which, as we have shown, can be considered as analogous to a ferromagnet) is dominating the physics of the system, at very low temperatures.

Also, for the out-of-equilibrium dynamics of the Edwards-Anderson model the RSB picture predicts a continuous violation of the FDT, associated to the existence of many ergodic components. \cite{Cugliandolo1993,Franz1994,Crisanti2003} Evidences of this have been found not only in numerical simulations of the 3D EAB model \cite{Barrat1998,Ricci2003} but also in recent experiments,\cite{Ocio2002} which lends strong support for the RSB picture.  But it has also been found \cite{Roma2007b} that when the system is separated into RL and FL the physical behavior of these components is very different from the one observed for the whole system. The FL curve is in perfect agreement with the FDT (as happens for a paramagnet) whereas for the RL the behavior found is similar to a coarsening process, which is typical of ferromagnetic materials. It is the combination of these two behaviors that gives rise to a violation of the FDT similar to what is predicted by the RSB picture.

It must be said that the idea that there is a special component of the system that is responsible for the singular behavior of spin glasses at low temperature is not new. \cite{Binder1986} In particular, an intuitive picture, based on the existence of ferromagnetic clusters, has been proposed to explain several experimental results. \cite{Mydosh1993} It postulates that the size of the clusters is inversely proportional to the temperature. More specifically, it assumes that the ferromagnetic clusters are composed by those spins joins by bonds that satisfy $J_{i,j}> T$. As a consequence, there appears a critical temperature at which the largest cluster percolates, resulting in the divergence of a correlation length. This picture, however, is essentially different to the one we propose in this  paper. In our case, the ferromagnetic-like clusters which form the backbone are only weakly related to the strength of its bonds, and depend instead on the structure of the GS and the first excited levels of the system. Another important difference is that the clusters are not independent, because there is an effective interaction between them, given by their embedding in a structure analogous to an excited phase.

%.....................................................................
\section{\label{Conc}CONCLUSIONS}

In this work we have carried out a detailed analysis of the properties and topology of the backbone of the Edwards-Anderson model with a bimodal distribution of bonds. The backbone is characterized by the RL, the structure formed by the bonds that do not change their condition in the different GS configurations of the system, and by the S spins, the set of spins which maintain the same relative orientation in these same configurations. We find that whereas in the thermodynamic limit there is a strong evidence that the backbone percolates in 3D, in 2D our results indicate that the most probable scenario is that this structure does not percolate. The results are consistent with the fact that only for the 3D EAB there is a positive critical temperature.  We also find that the frustration present in the RL is much smaller than the frustration of its complement, the FL. This leads to the conjecture that, at least at low enough temperatures, the RL and the FL could share many of the properties of a ferromagnet and a paramagnet, respectively. This conjecture has also been suggested in other contexts. \cite{Roma2006,Roma2007b,Roma2010}

In this paper the concept of RL is crucially dependent on the degeneracy of the GS of the system, and therefore applies only to models where the bonds can take only discrete values. However, as mentioned in the previous section, the idea of the separation of the system into two components with very different properties can be generalized to system with continuous distribution of bonds. We argue that the study of the separate contribution of these components to the different observables could lead to a deeper understanding of the low temperature phase of disordered and frustrated systems. In this sense, further work is being carried out to understand the properties of the backbone of continuous systems, as well as its influence on the probability distribution functions of the spin and link overlap and on the emergence of rare clusters in the paramagnetic phase of spin glasses (Griffiths singularities). \cite{Griffiths1969}

%.....................................................................
\begin{acknowledgments}
F. Rom\'a, A.J. Ramirez-Pastor and F. Nieto acknowledge support from CONICET (Argentina) under Project PIP112-200801-01332 and the National Agency of Scientific and Technological Promotion (Argentina) under Projects 33328 PICT-2005 and 2185 PICT-2007.  E. E. Vogel acknowledges partial support from the following Chilean agencies: Fondecyt (contract 1100156), Millennium Scientific Nucleus "Basic and Applied Magnetism" P06-022-F, and Center for the Development of Nanoscience and Nanotechnology Cedenna (Basal Funding program of Conicyt).  We thank S. Bustingorry, P.M. Gleiser, D. Dom\'{\i}nguez, L. F. Cugliandolo and J. Kurchan for fruitful discussions.

\end{acknowledgments}
%.....................................................................


\begin{thebibliography}{99}
\bibitem{Nota2}{We have checked that the results presented here for 2D and 3D systems agree, within statistical errors, with those corresponding to samples of lattice size $L>4$ where ferromagnetic and antiferromagnetic bods are chosen with $1/2$ probability.}
\bibitem{Nota1} {It is important to highlight that while in a model of random percolation
the concentration of bonds is imposed externally, the fraction of rigid bonds can not be chosen because the RL is a structure that emerges from a specified protocol defined in the GS.}
\bibitem{Nota3} Because the RL can be calculated for small lattices sizes and a limited number of samples, it is very difficult to use a standard procedure to extract the percolation exponents.  To overcome this problem we have chosen to fit the data using the Eq.~(\ref{scaling}), which gives good results for the random bond percolation of similar lattice sizes.
\end{thebibliography}
\end{document}